\documentclass[11pt]{article} 
\usepackage{xcolor}
\newcommand{\sop}[1]{\hat{\hat{#1}}}   
\usepackage{multirow}
  \usepackage{booktabs} 
\usepackage{multibib}
\usepackage[labelfont=bf]{caption}
\usepackage{booktabs}
\usepackage{graphicx}
\usepackage{authblk}
\usepackage[utf8]{inputenc}
\usepackage[a4paper, margin=1in]{geometry} 
\usepackage{caption}
\usepackage[version=4]{mhchem}
\usepackage{amsmath, amssymb, mathtools} 
\usepackage{dsfont} 
\usepackage{physics}
\usepackage{chemformula}
 
\usepackage{chemfig}
\usepackage{graphicx} 
\usepackage{float} 
\usepackage{subcaption} 
\usepackage{authblk} 
\usepackage{hyperref} 
\hypersetup{
    colorlinks=true,
    urlcolor=teal,
    linkcolor=orange,
    citecolor=olive
}
\usepackage{cite}

\usepackage{lineno} 
\usepackage{outlines}  
\usepackage{array}

\author[1,3,$*$]{Hadi Zadeh-Haghighi}
\author[2,3,$*$]{Gabriel E. Bertolesi}
\author[2,3]{Sarah McFarlane}
\author[1,3]{Christoph Simon}

\affil[1]{Department of Physics and Astronomy, Institute for Quantum Science and Technology, University of Calgary, Calgary, AB, Canada T2N 1N4}
\affil[2]{Department of Cell Biology and Anatomy, Alberta Children's Hospital Research Institute, University of Calgary, Calgary, Alberta, Canada}
\affil[3]{Hotchkiss Brain Institute, University of Calgary, Calgary, Alberta, Canada}

\affil[$*$]{These authors contributed equally to this work.}

\title{Magnetosensitivity of amphibian morphological pigmentation is light- and eye-dependent and consistent with the radical pair mechanism}
\date{\small Correspondence:
\{\href{mailto:hadi.zadehhaghighi@ucalgary.ca}{hadi.zadehhaghighi},
\href{mailto:gbertole@ucalgary.ca}{gbertole},
\href{mailto:smcfarla@ucalgary.ca}{smcfarla},
\href{mailto:csimo@ucalgary.ca}{csimo}\}@ucalgary.ca}

\date{\small Correspondence: \{hadi.zadehhaghighi, gbertole, smcfarla, csimo\}@ucalgary.ca}

\begin{document}

\maketitle

% \linenumbers 

\begin{abstract}
 
Weak magnetic fields influence a wide range of biological processes, yet the underlying mechanisms are poorly understood. The radical pair mechanism (RPM), which involves quantum spin dynamics, is a leading hypothesis. Here we show that weak magnetic fields modulate morphological pigmentation---specifically, the number of perioptic melanophores---in \textit{Xenopus laevis} tadpoles in a field-strength-dependent manner. The response is light- and eye-dependent. The observed field-strength dependence is quantitatively consistent with a radical pair model. These properties are reminiscent of the light-dependent magnetoreception that is thought to operate in migratory birds, and establish amphibian pigmentation as a tractable vertebrate system for the study of radical-pair quantum biology.
\end{abstract}
 
\noindent\textbf{Significance statement.} Whether weak magnetic fields can influence vertebrate physiology through quantum spin dynamics has been difficult to test in a living animal. We show that weak magnetic fields alter the number of perioptic melanophores in \textit{Xenopus laevis} tadpoles through a light- and eye-dependent mechanism whose field-strength dependence is quantitatively consistent with a radical-pair mechanism. Amphibian pigmentation thus offers a genetically tractable vertebrate system in which radical-pair quantum biology can be tested against whole-organism physiology. 
 
\section{Introduction}
 
A wide range of biological processes respond to weak magnetic fields (WMFs) of a
few millitesla or less, even though the associated interaction energies lie
orders of magnitude below thermal noise at physiological temperatures. These
effects---observed in avian magnetoreception~\cite{Cochran2004,Lisowski2026},
\textit{Drosophila} magnetosensitivity~\cite{Bradlaugh_2023}, stem-cell
behavior~\cite{Rishabh2026}, neurogenesis~\cite{Zhang2021}, axon
growth~\cite{Dufor2019}, cellular autofluorescence~\cite{Ikeya2021}, and
microtubule assembly~\cite{zadehSA2026}---resist explanation by classical
biochemistry and instead implicate the quantum dynamics of electron and nuclear
spins~\cite{ZadehHaghighi2022}.
 
A leading explanation for these phenomena is the radical pair mechanism (RPM)~\cite{ZadehHaghighi2022,Hore2016}. Here, pairs of radicals are generated whose unpaired electron spins occupy either a singlet state---antiparallel, entangled, and of total spin zero---or a triplet state of total spin one. Because spin--orbit coupling is weak in these light-atom radicals, electron spin is conserved to a good approximation during the reaction, so singlet and triplet pairs give rise to distinct products. Interactions with nearby nuclear spins, together with any external magnetic field, drive coherent oscillations between the two states~\cite{Steiner1989}; the rate of this interconversion, and hence the yield of each spin-selective product channel, depends on the field strength, providing a plausible biophysical basis for magnetosensitivity. In particular, a radical pair is suggested to form when blue light photoexcites the flavoprotein cryptochrome. This pathway is thought to underlie a retinal magnetic sense in migratory birds~\cite{Mouritsen2004}, where cryptochrome~4 (CRY4) is the leading candidate; purified CRY4 has been shown to host such magnetically sensitive radical pairs \textit{in vitro}~\cite{XuCRY2021}.

Amphibian pigmentation offers a tractable vertebrate readout for probing such a mechanism. In \textit{Xenopus laevis}, WMFs were reported decades ago to alter background adaptation~\cite{Leucht1987} and to affect isolated tail-fin melanophores \textit{in vitro}~\cite{Leucht1987b}. In a separate, gravity-referenced assay, \textit{Xenopus} larvae showed directional selectivity to the \textit{inclination} of a weak field---the same signature as the avian inclination compass---which was abolished by optic-nerve transection, implicating the peripheral visual system~\cite{Leucht1990}. These studies established the phenomenon but did not systematically characterise its field-strength dependence or propose a mechanism. They also left open whether light \textit{itself} is required for the eye-mediated effect, whether the response reflects a dedicated sensory pathway, and whether it extends to \textit{morphological} pigmentation: the slower, eye-controlled change in perioptic melanophore \textit{number}~\cite{Bertolesi2016,AtkinsonLeadbeater2024}, which is quantifiable and distinct from the rapid pigment movements examined earlier.
 
We therefore examined morphological pigmentation in \textit{Xenopus laevis}, which retains \textit{cry4}~\cite{Takeuchi2014}---a gene lost in mammals~\cite{Haug2015} and the leading candidate magnetoreceptor in migratory birds~\cite{PinzonRodriguez2018,XuCRY2021}. Fields of 0.25--1,mT increased pigmentation relative to geomagnetic-field (GMF) controls in a field-strength-dependent manner. The response was eye-dependent, required light---consistent with a dedicated, light-dependent sensor, and paralleling the radical-pair magnetoreception proposed in migratory birds.
 
To assess whether a radical-pair mechanism could account for this WMF
sensitivity, we performed spin-dynamics simulations for two radical pairs: one
parameterized with hyperfine coupling constants from avian CRY4---the
biologically motivated candidate---and, because the responsible pair is unknown,
a generic pair not tied to any specific molecule. Both quantitatively reproduce
the measured field-strength dependence with physiologically plausible parameters.
 
Together, these results characterise a light- and eye-dependent magnetic-field effect on \textit{Xenopus} pigmentation and show its field-strength dependence to be consistent with a radical-pair mechanism, while leaving open the molecular identity of the pair and the directional properties of the response. The remainder of the paper presents the experimental characterisation and the spin-dynamics modelling used to assess a radical-pair origin. We then discuss the implications for cryptochrome-based magnetoreception and for \textit{Xenopus} as a tractable system for quantum-biology studies.
 
\section{Results}
\subsection{Magnetic fields modulate light-dependent morphological pigmentation}
\begin{figure}[H]
\centering
\includegraphics[width=1\textwidth]{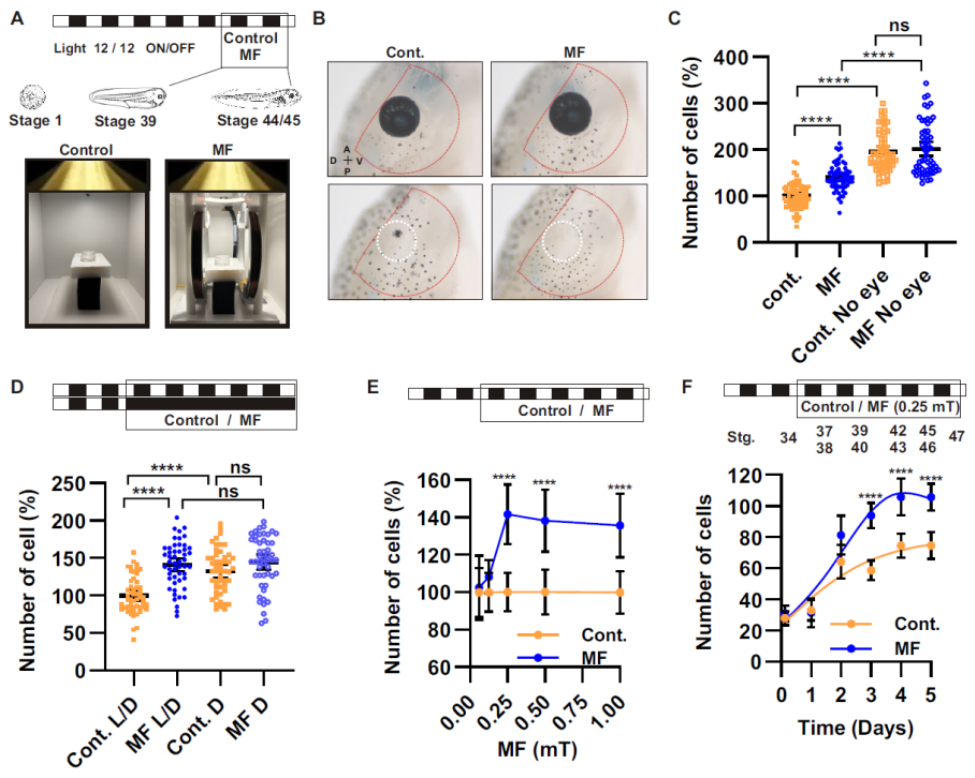}
\caption{\textbf{Low magnetic fields increase morphological pigmentation.} \textbf{A)} Schematic of the experimental protocol used. Stage 35 or 39 embryos were raised on a white surface with light/dark cycles and exposed to a magnetic field (MF) generated with a Ferronato\textregistered{} BH230HP-A Helmholtz coil apparatus (see also Supplementary Fig.~\ref{fig:figureS1}). Note the dishes containing tadpoles in the middle of the coil. \textbf{B, C)} Lateral views of the perioptic region used to score morphological pigmentation in control and enucleated tadpoles exposed to 0.5\,mT MF for two days starting from stage 39 until stage 43/44 \textbf{(B)}, with perioptic melanophore number per eye shown in \textbf{(C)}. Each dot represents the number of perioptic melanophores normalized to the corresponding control (100\%). Data show the mean $\pm$ 95\% confidence interval from three independent experiments (N~=~3; $n$~=~28--34 tadpoles per group). \textbf{D)} Morphological pigmentation in tadpoles kept in constant darkness (D) from stage 35 versus those exposed to a circadian light/dark cycle (L/D) with or without MF exposure (0.75\,mT). \textbf{E)} Quantification of perioptic melanophores in tadpoles exposed to various MF strengths. \textbf{F)} Daily measurement of perioptic melanophore pigmentation in tadpoles exposed to 0.25\,mT from stage 35; a statistically significant increase was observed at stage 43/44, two days after the onset of visual function (stage 39). Pigmentation data across different hatches were normalized and expressed as a percentage relative to their sister experimental controls. Statistical significance was assessed using multiple ANOVA followed by Tukey's post hoc test. ns, not significant; *\textit{p}~$<$~0.05; **\textit{p}~$<$~0.01; ***\textit{p}~$<$~0.001; ****\textit{p}~$<$~0.0001.}
\label{fig:figure1}
\end{figure}

We investigated the direct effects of WMF on \textit{Xenopus laevis} tadpoles while carefully controlling for additional factors that could influence skin pigmentation, with all perioptic melanophore counts scored blind to the experimental condition. Embryos were placed in a horizontally oriented MF setup (Fig.~\ref{fig:figure1}A and Supplementary Fig.~\ref{fig:figureS1}) that maintained stable temperatures ($\pm$0.2\,$^{\circ}$C difference between control and experimental conditions; Supplementary Fig.~\ref{fig:figureS2}) and controlled light conditions (12\,h light ON / 12\,h light OFF). Pigmentation was assessed during the mid-light phase (Zeitgeber time +5 to +7) to account for possible circadian variations in skin coloration, and on a white surface. We chose to investigate skin pigmentation changes on a white background, as subtle MF effects are masked by the much larger melanophore increase induced by a black background. WMF-exposed specimens displayed a significantly increased number of melanophores in the perioptic region (Fig.~\ref{fig:figure1}B,C), a response known to be regulated by eye photosensitivity~\cite{Bertolesi2016,AtkinsonLeadbeater2024}; with the eye intact, WMF (0.5\,mT) increased perioptic melanophores by 37\% ($g=0.90$, $p<0.0001$; $N=3$, $n=28$--$34$ per group). Enucleation at stage 39 raised the baseline melanophore number $\sim$1.9-fold ($g=2.04$, $p<0.0001$) and abolished the detection of the magnetic-field response ($g=0.00$, $p=0.997$), suggesting that the effect requires an intact eye (Fig.~\ref{fig:figure1}B,C). Exposure to WMF beginning at developmental stage 34/35---when organogenesis is advanced and melanophores are already visible in the skin---consistently produced a large increase in morphological pigmentation under a light/dark cycle ($+$75\%; $g=2.22$, $p<0.0001$; 0.75\,mT; Fig.~\ref{fig:figure1}D). In constant darkness the baseline was itself substantially elevated ($+37\%$ relative to the light control; $g=0.95$, $p<0.0001$), yet WMF produced no further increase ($+4\%$; $g=0.13$, $p=0.51$; Fig.~\ref{fig:figure1}D). Darkness and WMF thus drive the same phenotype non-additively: WMF increased pigmentation only while light-dependent repression was operative (light/dark cycle) and had no effect once that repression was relieved by constant darkness. Such occlusion is expected if the two stimuli converge on a single mechanism rather than acting independently, indicating that the magnetic-field effect requires light and that WMF relieves the same light-dependent repression that darkness removes. We found previously that light acting through the eye was essential to repress melanophore differentiation in the perioptic region~\cite{AtkinsonLeadbeater2024}, suggesting involvement of the eye in the magnetoreceptive regulation of skin pigmentation and implying that magnetic stimulation counteracts the repression mechanism. Consistent with a retinal contribution, the effect was not detectable until stage 43/44 (3 days at room temperature; Fig.~\ref{fig:figure1}F), after the onset of retinal functionality~\cite{Holt1984}. The response was dependent on magnetic flux density: 0.25\,mT was the lowest field strength that reliably induced measurable pigmentation changes (Fig.~\ref{fig:figure1}E), whereas 0.125\,mT (roughly twice the geomagnetic field in Calgary, Alberta; 0.055\,mT) and 0.06\,mT failed to produce pigmentation levels significantly different from controls at that time point (Fig.~\ref{fig:figure1}E). Importantly, embryos exposed to WMF for five days starting at stage 35 displayed gross morphology comparable to controls (Supplementary Fig.~\ref{fig:figureS3}), indicating that normal development is unaffected by WMF.
 
\subsection{Radical pair model for magnetic field effect on perioptic melanophores}
 
Across 0.125--1\,mT the perioptic pigmentation response rose with field strength (Fig.~\ref{fig:figure1}E). The cryptochrome photocycle generates several candidate spin-correlated radical pairs (See Methods); to test whether this field-strength dependence is consistent with a radical-pair mechanism, we simulated the flavin-based $[\text{FAD}^{\bullet-}\text{---}\text{TrpH}^{\bullet+}]$ pair as a biologically motivated representative and a generic $[\text{H,H}^{\bullet}\text{---}\text{H,H}^{\bullet}]$ reference pair. Each is treated as a singlet-born spin-correlated radical pair whose singlet$\leftrightarrow$triplet interconversion, and thus the final yield, are driven by the Zeeman and isotropic hyperfine interactions, with the magnetic-field effect given by the field dependence of the time-integrated triplet yield scaled by an empirical amplification factor ($\alpha$). This factor represents the biological gain that links the small, spin-dependent change in triplet yield to the macroscopic pigmentation response; some such amplification is generically required in radical-pair magnetoreception, where the primary spin-chemical signal is usually small~\cite{Kattnig_2016,Player2021}, and we treat it as a single free scaling parameter rather than modelling its origin. The singlet-recombination and triplet/escape rates ($k_S$, $k_T$) were fitted to the field-strength dependence, with the hyperfine couplings fixed at literature values (see Table~\ref{tab:hfcc}) for the flavin-based pair and fitted (within $|a|\in[0.4,1.0]$\,mT) for the generic pair (see Methods).
 
\begin{figure}[H]
    \centering
    \includegraphics[width=0.85\textwidth]{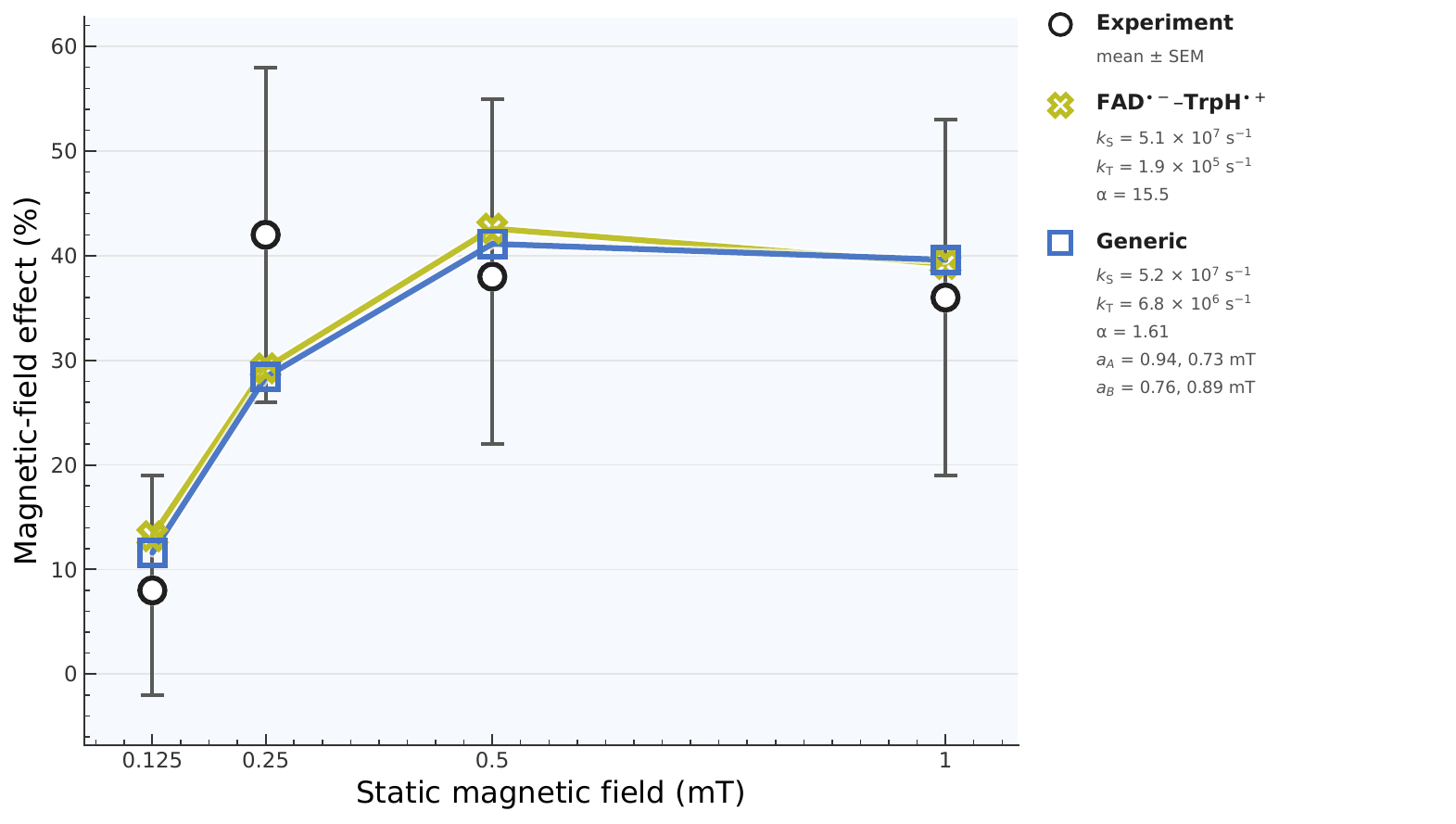}
    \caption{\textbf{Modeling magnetic field effects on perioptic melanophores in \textit{Xenopus laevis}.} \emph{Experimental data} (open circles, bars): the magnetic-field effect (MFE) on the cellular pigmentation readout--- the percent change of the magnetically treated sample relative to the unexposed control --- measured at four static fields, $B_0 = 0.125, 0.25, 0.5, 1.0$\,mT (Fig.~\ref{fig:figure1}E); bars give the measured range, and the $0.05$\,mT point is used as the field reference. \emph{Model:} each is a singlet-born spin-correlated radical pair evolving under the Zeeman + isotropic-hyperfine spin Hamiltonian with Haberkorn singlet recombination ($k_\mathrm{S}$), triplet/escape ($k_\mathrm{T}$) and fixed isotropic spin relaxation ($k_\mathrm{rel}=10^{6}\,\mathrm{s^{-1}}$); the MFE is the field dependence of the time-integrated triplet yield, $B_\mathrm{ref}=0.05$\,mT. \emph{Modeled pairs and isotropic hyperfine couplings} ($a$ in mT; $\mathrm{N}={}^{14}$N, $I=1$; $\mathrm{H}={}^{1}$H, $I=\tfrac12$): $\mathrm{FAD}^{\bullet-}$--TrpH$^{\bullet+}$ [FAD$^{\bullet-}$: N $+0.57$, H $+0.42$] and a generic $(\mathrm{H,H})$--$(\mathrm{H,H})$ pair with four proton couplings fitted under $|a|\in[0.4,1]$\,mT. For each pair the best ($\chi^2$-ranked) data-consistent sets are shown.  \emph{Amplification required:} the two models demand very
  different amplification to reach the measured effect. The
  fixed-coupling $\mathrm{FAD}^{\bullet-}$--TrpH$^{\bullet+}$ pair
  needs $\alpha\approx15$ for its best fit ($\chi^2\approx0.99$;
  data-consistent solutions exist only up to $\alpha\approx100$),
  whereas the generic $(\mathrm{H,H})$--$(\mathrm{H,H})$ pair
  reproduces the data equally well or better ($\chi^2\approx0.92$)
  with $\alpha\approx1.6$ --- close to the unamplified limit and
  roughly an order of magnitude less amplification.}
    \label{fig:figure2}
\end{figure}

Figure~\ref{fig:figure2} shows the measured effect at the four relevant fields ($0.125$--$1.0$\,mT) alongside the corresponding fits. The effect climbs sharply from 0.125 to 0.25\,mT and then levels off near $40\%$ out to 1.0\,mT---a field dependence that both pairs reproduce; for the flavin-based pair the fit gives fast recombination ($k_S\approx5\times10^{7}$\,s$^{-1}$) and slow triplet/escape ($k_T\approx2\times10^{5}$\,s$^{-1}$). Since each pair fits well across a wide, mutually overlapping range of ($k_S$, $k_T$), the two pairs cannot be told apart and the present data place no meaningful bound on the individual rate constants (Supplementary Fig.~\ref{fig:figureS4}). The two models do, however, require markedly different amplification to match the observed effect. The fixed-coupling $\mathrm{FAD}^{\bullet-}$--TrpH$^{\bullet+}$ pair attains its best fit at $\alpha\approx15$ ($\chi^2\approx0.99$), with data-consistent solutions extending only to $\alpha\approx100$, whereas the generic $(\mathrm{H,H})$--$(\mathrm{H,H})$ pair matches the data equally well or better ($\chi^2\approx0.92$) at $\alpha\approx1.6$---near the unamplified limit and roughly an order of magnitude less amplification.
 
\section{Discussion}
 
The results presented here show that weak magnetic fields modulate morphological pigmentation in \textit{Xenopus laevis} through a light- and eye-dependent mechanism (Fig.~\ref{fig:figure1}). The field-strength dependence is closely matched by RPM-based spin-dynamics simulations, consistent with a cryptochrome radical-pair mechanism as a feasible biophysical basis for the effect (Fig.~\ref{fig:figure2}). To our knowledge this is the first evidence that \textit{morphological} (melanophore-number) pigmentation---as distinct from the rapid background-adaptation response examined decades ago~\cite{Leucht1987}---is magnetosensitive. The morphological effect required somewhat stronger fields than the earlier rapid responses: those were elicited at geomagnetic-scale intensities ($\leq$0.08\,mT)~\cite{Leucht1987}, whereas the melanophore-number response emerged only well above the geomagnetic value (Fig.~\ref{fig:figure1}E). We observe a clear dependence on light itself, not merely on an intact optical system---a light-dependence not reported in the earlier eye-mediated work~\cite{Leucht1990}. These properties parallel the light-dependent radical-pair mechanism proposed for magnetoreception in migratory birds.
 
Previous work from our laboratory showed that perioptic melanophore
differentiation is repressed under light and de-repressed by darkness, eye
enucleation, optic-nerve transection, or pharmacological disruption of
photoreceptor--ganglion-cell signalling~\cite{Bertolesi2016,AtkinsonLeadbeater2024},
implicating an inhibitory tone from retinal photoreceptors. The light- and
eye-dependent magnetic-field induction reported here joins these perturbations
in relieving the repression, and their convergence on a common phenotype points
to a retinal photoreceptive pathway. CRY4 is a plausible candidate mediator of
WMF effects: it localizes to the outer segments of retinal cone
photoreceptors~\cite{Gnther2018} and has a proposed role in avian
magnetoreception~\cite{XuCRY2021}. If CRY4 is the light sensor maintaining this
repression, the darkness result follows naturally: with the repression already
lifted, magnetic-field exposure adds nothing, as expected if the field acts
through the same pathway---plausibly via CRY4 radical pairs. Two observations
argue that this reflects the sensory pathway rather than a trivial ceiling
effect: in constant darkness the baseline rose but remained below the eyeless
level yet still showed no magnetic-field increase, and the eye-intact response is
itself graded with field strength rather than saturated (Fig.~\ref{fig:figure1}D,E).
Because we assayed at fixed Zeitgeber times (ZT 6 and ZT 18) and abolished the
response by removing light rather than by perturbing the clock, we interpret the
effect as light-driven rather than clock-gated, although a clock-specific
manipulation remains to be performed. Unlike the opsins implicated in
physiological pigmentation responses (pinopsin, melanopsin)~\cite{Bertolesi2016,Bertolesi2022,Heshami2026},
CRY4 would be the first photosensor implicated in morphological
pigmentation---a distinct photosensory pathway on a slower, structural timescale.
That our melanophore-number effect requires both light and the eye is the
signature expected of a retinally mediated response, potentially triggered by a
cryptochrome-based radical pair.
 
Because the radical pair underlying the response is unidentified, we used two models: (i)~a biologically motivated one with HFCCs from avian retinal CRY4a~\cite{Benjamin_2025,Deviers2022} for the $[\text{FAD}^{\bullet-}\text{---}\text{TrpH}^{\bullet+}]$ pair of the cryptochrome photocycle (Fig.~\ref{fig:figure3}) and (ii)~a generic pair ($[\text{HxH}^{\bullet}\text{---}\text{HxH}^{\bullet}]$) with isotropic HFCCs fitted to the field-strength dependence. The data comprise only four field points, and they do not constrain either pair's recombination and escape rates: both pairs reproduce the field-strength dependence over broad, overlapping regions of ($k_S$, $k_T$). For the generic pair, hyperfine couplings restricted \textit{a priori} to the physically reasonable range for organic radicals ($|a|\in[0.4,1.0]$~mT)~\cite{Gerson_2003} suffice to reproduce the data. We read the modelling as establishing that the RPM is \textit{feasible}---a radical pair of this class can produce the measured field dependence---rather than as evidence for any specific pair. Indeed, the unbiased generic pair fits at least as well as the flavin-based pair ($\chi^2\approx0.92$ vs.\ $0.99$) while requiring far less amplification. The amplification factor needed to reach the macroscopic effect is here modest---close to unity ($\alpha\approx1.6$) for the generic pair and of order ten ($\alpha\approx15$) for the flavin--tryptophan pair. This is a far less demanding regime than the geomagnetic-strength models, where the spin-chemical signal to be amplified is tiny and that amplification is the central difficulty~\cite{Kattnig_2016,Player2021}; the singlet--triplet modulation at higher field is much larger, so our data do not bear on how a geomagnetic-strength field could be amplified. Despite requiring roughly an order of magnitude more amplification than the generic pair, the flavin--tryptophan pair is among the most studied in this context~\cite{Wong2021,Atkins2019,Deviers2022}, making a cryptochrome radical pair a reasonable candidate class; the link to CRY4 itself remains correlative absent loss-of-function, radiofrequency-disruption, or action-spectrum data.
 
Although the data cannot identify the radical pair, the mechanism makes qualitative predictions that are independent of the pair's identity and directly testable. The most distinctive is non-monotonicity: the data-consistent solutions for both pairs rise to a maximum and then reverse sign at higher field (Supplementary Fig.~\ref{fig:figureS4})---a turnover that is a hallmark of radical-pair spin dynamics. Because the rate constants are unconstrained, the precise turnover is not pinned down (the maximum lies below $\sim$1\,mT and the sign reversal in the low-millitesla range, roughly $1.5$--$3$\,mT depending on the pair and parameters), but its qualitative existence is robust. Note that going to higher fields will require managing resistive heating from the Helmholtz coils. A second identity-independent diagnostic is radiofrequency sensitivity: a weak resonant radiofrequency field should disrupt the response. Identifying \textit{which} pair is involved would instead require identity-sensitive measurements---the frequency profile of that radiofrequency response, which is set by the radicals' hyperfine couplings; \textit{in vitro} transient-absorption spectroscopy of the candidate radicals; or, since we have shown only that light is required, a test of whether the effect specifically needs short-wavelength (blue/ultraviolet) light matching flavin absorption, the signature of a cryptochrome or potentially other flavoprotein.
 
A direct test of CRY4's necessity will require loss-of-function. \textit{Xenopus laevis} is amenable to CRISPR--Cas9 editing~\cite{Blitz2021,Heshami2026}, so \textit{cry4} knockouts---testing whether the magnetic-field response is abolished while other phenotypes remains intact---are a clear next step toward causality. 
 
A further open question concerns the geometry of the response. We varied only field strength, holding its direction fixed, and did not manipulate field orientation or polarity; our spin Hamiltonian is correspondingly isotropic, predicting a dependence on field magnitude alone, and the measured response is well described on this basis. The data therefore neither demonstrate nor exclude a directional component of the pigmentation effect. Earlier behavioural work in \textit{Xenopus} larvae sharpens this question. There, gravity-oriented body-axis posture tracked the \textit{inclination} of a near-geomagnetic field, the response reversing in sign as the dip was changed among $0^{\circ}$, $35^{\circ}$ and $90^{\circ}$~\cite{Leucht1990}. That study also reported a heading-dependent north--south asymmetry that Leucht read as field polarity. Because a dipping field can produce such an asymmetry through inclination geometry alone, we regard the polar interpretation as unresolved rather than established. Testing polarity, however, in our experiment is complicated by the tadpoles' free movement, which leaves the field orientation relative to each animal uncontrolled over the days-long exposure.
 
Together, these results show that a weak-magnetic-field response in \textit{Xenopus} pigmentation is light- and eye-dependent and quantitatively consistent with a cryptochrome radical pair mechanism, strengthening the radical-pair hypothesis in a vertebrate system while leaving the definitive test of CRY4's necessity to future loss-of-function work. Its significance extends beyond amphibian pigmentation: if a conserved photoreceptor serves as a magnetic sensor in \textit{Xenopus}, analogous mechanisms may operate in the retinae of migratory birds and other light-exposed tissues. The quantitative agreement provides rare \textit{in vivo} evidence that coherent spin dynamics---usually confined to isolated molecules in the laboratory---can survive the thermal and molecular complexity of living tissue to shape a macroscopic outcome, establishing \textit{Xenopus laevis} as a tractable vertebrate system in which the predictions of quantum mechanics can be tested against whole-organism biology.

\section*{Methods}
\subsection*{Magnetic field setup}
Magnetic fields were generated using Ferronato\textregistered{} BH230HP-A high-power, single-axis Helmholtz coils (Bartington Instruments) with a nominal diameter of 230 mm. This configuration comprises two identical circular coils positioned coaxially at a separation equal to their radius, yielding a uniform magnetic field within the central volume. Such uniformity is critical for experiments investigating magnetic field effects on radical pairs, as it reduces spatial gradients and ensures reproducible exposure across samples. 
The experimental apparatus, together with the simulated magnetic field distribution, is depicted in Supplementary Fig.~\ref{fig:figureS1}. According to the manufacturer, the field homogeneity is within $\pm$0.25\% over a 90 mm diameter and length, while $\pm$5\% uniformity extends over the entire blue area depicted in Supplementary Materials, beyond 100 mm in diameter and length.
 
\subsection*{Temperature control}
Embryos were maintained in white boxes under magnetic-field or control conditions with uniform overhead illumination ($\sim$1000 lux; 12 h light / 12 h dark) and reared against a white background. We restricted measurements to fields of 1 mT or below: above this intensity, Joule heating from the resistive coils produced temperature differences between the magnetic-field and control boxes that could confound the comparison. At 1 mT and below, this difference remained under 0.2
$^{\circ}$C (Supplementary Fig.~\ref{fig:figureS2}).
 
\subsection*{Embryos and magnetic field}
The Animal Care and Use Committee, University of Calgary, approved procedures involving frogs and embryos (AC21-0148; signed by Dr. Derrick Rancourt). Embryos were obtained by induced egg production from chorionic gonadotrophin (Intervet Canada Ltd.) injected females and in vitro fertilization according to the standard procedures (see protocols at Xenbase, http://www.xenbase.org). Embryos were maintained at 16 $^{\circ}$C in Marc's modified Ringer's (MMR) solution (100mM NaCl, 2mM KCl, 2mM CaCl$_2$, 1mM MgCl$_2$, 5mM HEPES pH 7.4) until stage 43/44 (approximately 1 week) and staged according to Nieuwkoop and Faber on Xenbase (www.xenbase.org). Of note, embryos at this stage cannot be sexed. The embryos were reared under light cycles of 12 h ON / 12 h OFF (light = 1000 lux or approximately $1.5\times10^{-4}$\,W\,cm$^{-2}$) on a white background, and exposed to a horizontal magnetic field generated by the Helmholtz-coil apparatus described above (Supplementary Fig.~\ref{fig:figureS1}).
Field strength was selected per assay: 0.06--1.0\,mT for the field-strength series, 0.5\,mT for the enucleation experiment, and 0.75\,mT for the light/dark experiment. In the enucleation assay, embryos underwent bilateral eye removal at stage 39/40, recovered for 4\,h, and were then exposed---intact or enucleated---to the magnetic field for 48\,h until stage 44/45. In the light/dark assay, embryos were exposed from stage 35 for four days, with light animals maintained on the 12\,h ON / 12\,h OFF cycle and dark animals kept in a light-tight box. Within each assay, control and magnetic-field groups were processed in parallel, and all assays were quantified and analysed identically (perioptic melanophore counts scored blind to treatment; see \textit{Determination of morphological pigmentation} and \textit{Statistical analysis}).
 
\subsection*{Determination of morphological pigmentation (perioptic melanophore cell numbers)}
For pigment cell counting, cells were treated briefly with melatonin prior to being fixed to aggregate the melanosomes, which allowed for easy identification of individual melanophores as each melanophore contained a single aggregate of pigment. Fixed embryos were imaged around the eyes, and a region (1 eye diameter) was selected from each image in Adobe Photoshop, copied, and randomly arranged on a grid so that all melanophore counting was performed blind to the treatment conditions (see Fig.~\ref{fig:figure1}B).

\subsection*{Spin dynamics simulations}
 
\begin{figure}[H]
    \centering
    \includegraphics[width=1\textwidth]{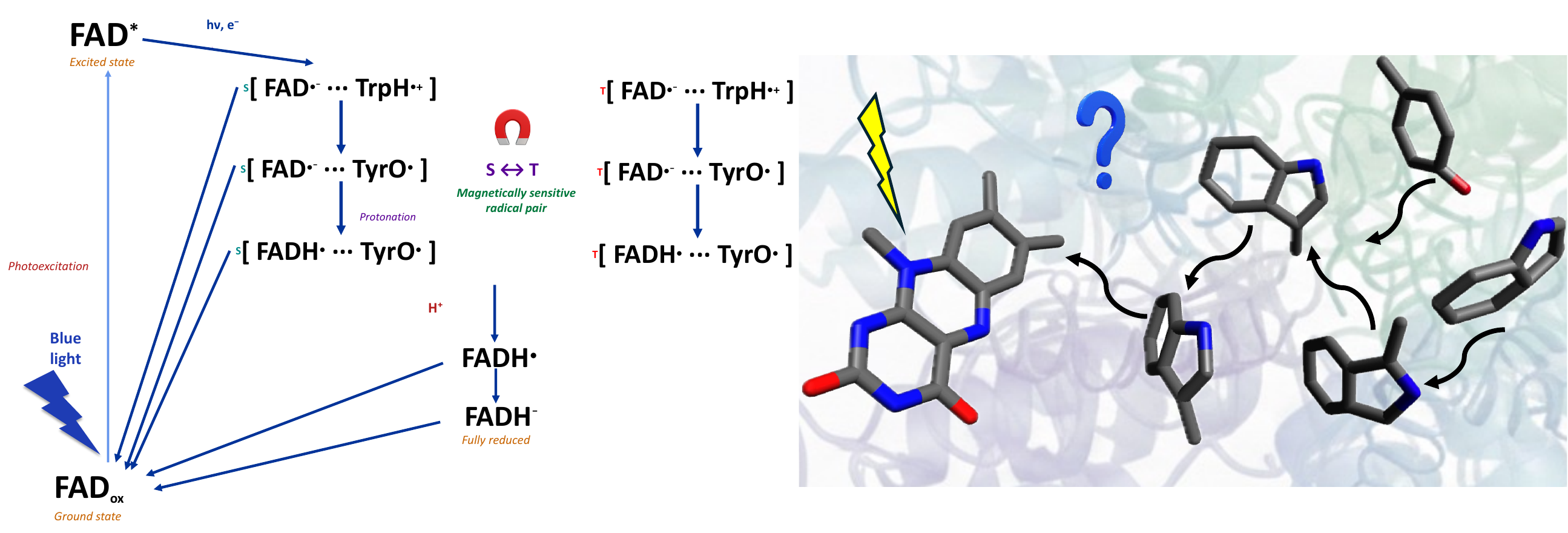}
    \caption{\textbf{Cryptochrome photocycle and candidate radical pairs.} \textbf{(Left)} Photocycle and spin kinetics: blue-light excitation of FAD, followed by sequential electron transfer and protonation, generates the candidate spin-correlated radical pairs $^{1,3}[\mathrm{FAD}^{\bullet-}\!\cdots\mathrm{TrpH}^{\bullet+}]$, $^{1,3}[\mathrm{FAD}^{\bullet-}\!\cdots\mathrm{TyrO}^{\bullet}]$ and $^{1,3}[\mathrm{FADH}^{\bullet}\!\cdots\mathrm{TyrO}^{\bullet}]$, whose singlet$\leftrightarrow$triplet interconversion is modulated by the static magnetic field; recombination and escape return the system to the $\mathrm{FAD_{ox}}$ ground state (approximate lifetimes indicated). The $[\mathrm{FAD}^{\bullet-}\!\cdots\mathrm{TrpH}^{\bullet+}]$ pair is the flavin-based pair modelled here (Fig.~\ref{fig:figure2}). \textbf{(Right)} Molecular view of the flavin and the candidate tryptophan/tyrosine electron-donor partners.}
    \label{fig:figure3}
\end{figure}

We modelled two spin-correlated radical pairs: a flavin-based pair $\mathrm{FAD}^{\bullet-}\!\cdots\mathrm{TrpH}^{\bullet+}$, of the type proposed in the cryptochrome photocycle (Fig.~\ref{fig:figure3}), and a generic $(\mathrm{H,H})$--$(\mathrm{H,H})$ reference pair.
 
Radical-pair spin dynamics were simulated in Liouville (superoperator) space for a system comprising the two unpaired electron spins together with their hyperfine-coupled nuclei. The spin Hamiltonian, in angular-frequency units, contained only the electron Zeeman and isotropic hyperfine interactions,
\begin{equation}
\hat{\mathcal{H}} = \gamma_e B_0\,\qty(\hat{S}_{A,z}+\hat{S}_{B,z}) + \sum_{i} a_{A,i}\,\gamma_e\,\hat{\boldsymbol{S}}_A\vdot\hat{\boldsymbol{I}}_{A,i} + \sum_{j} a_{B,j}\,\gamma_e\,\hat{\boldsymbol{S}}_B\vdot\hat{\boldsymbol{I}}_{B,j},
\label{eq:hamiltonian}
\end{equation}
where $\gamma_e = 1.76086\times10^{8}$\,rad\,s$^{-1}$\,mT$^{-1}$ is the electron gyromagnetic ratio (isotropic $g$-factor, identical for both radicals), $B_0$ is the static field applied along $z$, and the $a_{X,i}$ are isotropic hyperfine coupling constants (in mT). Nuclear Zeeman, inter-radical exchange and electron--electron dipolar couplings were neglected; because all retained interactions are isotropic, the predicted effect depends only on the magnitude of $B_0$, consistent with our intensity-only measurements, in which the field direction was held fixed and orientation dependence was not probed. Each radical was represented by two hyperfine-coupled nuclei of spin $\tfrac{1}{2}$ ($^{1}$H) or spin $1$ ($^{14}$N); for the flavin-based pair the isotropic couplings were taken from European robin CRY4~\cite{Benjamin_2025,Deviers2022} (see Table~\ref{tab:hfcc}), while for the generic model the four couplings were treated as free parameters within $|a|\in[0.4,1.0]$\,mT. 
 
The radical pair was initialised in the electronic singlet state with nuclei maximally mixed, $\hat{\rho}(0)=\hat{P}_S\otimes(\hat{\mathds{1}}_n/Z)$, where $\hat{\mathds{1}}_n$ is the nuclear identity of dimension $Z$. Its evolution was governed by a Liouville--von Neumann equation with Haberkorn spin-selective recombination and phenomenological spin relaxation,
\begin{equation}
\dv{\hat{\rho}(t)}{t} = -i\comm{\hat{\mathcal{H}}}{\hat{\rho}(t)} - \tfrac{1}{2}\acomm{\hat{\mathcal{K}}}{\hat{\rho}(t)} + \sum_{X=A,B}\sop{\mathcal{R}}_X\qty[\hat{\rho}(t)], \qquad \hat{\mathcal{K}} = k_S\hat{P}_S + k_T\hat{P}_T,
\label{eq:master}
\end{equation}
where $\hat{P}_S$ and $\hat{P}_T$ are the electronic singlet and triplet projectors and $k_S$, $k_T$ are the singlet and triplet decay rates. Relaxation was represented by Lindblad jump operators $\sqrt{k_{\mathrm{relax}}}\,\hat{S}_{X,\alpha}$ ($\alpha=x,y,z$) acting independently on each electron, with $k_{\mathrm{relax}}^{A}=k_{\mathrm{relax}}^{B}=10^{6}$\,s$^{-1}$~\cite{XuCRY2021}. The ultimate triplet yield was evaluated from the time-integrated density matrix $\bar{\hat{\rho}}=\int_0^\infty\hat{\rho}(t)\,\dd t=-\sop{\mathcal{L}}^{-1}\hat{\rho}(0)$, obtained by solving the associated sparse linear system rather than by explicit time propagation, as $\Phi_T = k_T\,\Tr[\hat{P}_T\bar{\hat{\rho}}]$. The magnetic-field effect was defined as the percentage change in $\Phi_T$ relative to a reference field $B_{\mathrm{ref}}=0.05$\,mT (approximately the local geomagnetic intensity) and mapped onto the measured pigmentation response by a single linear amplification factor $\alpha$~\cite{Kattnig_2016,Player2021}:
\begin{equation}
\mathrm{MFE}(B)=\alpha\,100\,[\,\Phi_\mathrm{T}(B)/\Phi_\mathrm{T}(B_\mathrm{ref})-1\,].   
\end{equation}
 
The decay rates $k_S$ and $k_T$ and $\alpha$ were fitted for the $\mathrm{FAD}^{\bullet-}\!\cdots\mathrm{TrpH}^{\bullet+}$ pair (hyperfine couplings fixed at literature values), and the four hyperfine couplings together with $k_S$ and $k_T$ for the generic pair ($k_S,k_T\in[10^{4},10^{8}]$\,s$^{-1}$), against the measured field-strength dependence using the Differential Evolution algorithm~\cite{Rocca2011}, with $k_{\mathrm{relax}}$ held fixed. Because only four field strengths were available against the fitted rate constants (and, for the generic pair, the hyperfine couplings), we enumerated the full set of data-consistent solutions for each pair rather than reporting a single fit (Supplementary Fig.~\ref{fig:figureS4}).
 
\subsection*{Statistical analysis}
Statistical analyses are described in the individual figure legends and in the method sections for each separate experimental assay. Statistical significance was assessed using multiple ANOVA followed by Tukey's post hoc test. In addition to significance testing, magnetic-field effects were quantified as standardised effect sizes (Hedges' $g$) with their standard errors (Supplementary Fig.~\ref{fig:figureS6}), and the per-field comparisons in the field-strength experiment were also corrected for multiple comparisons using the Holm procedure (reported in the Supplementary Figs.~\ref{fig:figureS5} \&~\ref{fig:figureS6}).

\section*{Data, code, and materials availability:}
All data and code needed to evaluate and reproduce the results in the paper are present in the paper and/or the Supplementary Materials.

\section*{Acknowledgments}
We thank Ryan Bui and Carrie Hehr for their excellent technical assistance, Rishabh for useful discussions, and GMW Associates for providing the FEMM simulation (Fig.~\ref{fig:figureS1}). 

\section*{Funding}
This work was supported by a Natural Sciences and Engineering Research Council of Canada (NSERC) Discovery Grant (S.M.), an NSERC Discovery Grant (C.S.), the Alliance quantum consortia grant 'Quantum Enhanced Sensing and Imaging' (QuEnSI ALLRP 578468-22) (C.S.), and a National Research Council (NRC) Quantum Sensing Challenge program grant (CSTIP QSP 022) (C.S.).

\section*{Conflict of interest }

The authors declare no conflict of interest.

\section*{Author contributions}
H.Z-H.: Writing—original draft, conceptualization, investigation, writing—review and editing, methodology, data curation, validation, supervision, formal analysis, software, project administration, and visualization. G.E.B.: Writing—original draft, conceptualization, investigation, writing—review and editing, methodology, data curation, validation, supervision, formal analysis, software, project administration, and visualization. S.M.: Writing—original draft, conceptualization, investigation, writing—review and editing, validation, supervision, formal analysis, project administration, resources, and funding acquisition. C.S.: Writing—original draft, conceptualization, investigation, writing—review and editing, validation, supervision, formal analysis, project administration, resources, and funding acquisition.

\bibliographystyle{unsrt}  
\bibliography{Ref}

\clearpage
\section*{Supplementary Materials}
\renewcommand{\thefigure}{S\arabic{figure}}
\setcounter{figure}{0}
 \renewcommand{\thetable}{S\arabic{table}}
\setcounter{table}{0}
\begin{figure}[H]
\centering
\includegraphics[width=0.8\textwidth]{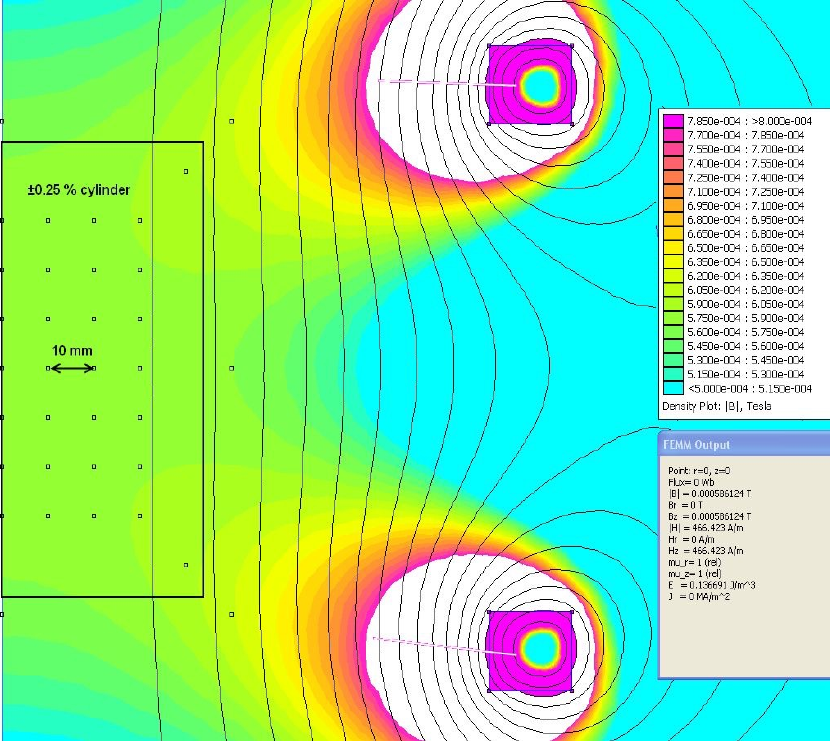} 
\caption{Simulated magnetic field distribution generated by the Helmholtz coils, provided courtesy of GMW Associates and reproduced with their written permission.}
\label{fig:figureS1}
\end{figure}
 
\begin{figure}[H]
\centering
\includegraphics[width=1\textwidth]{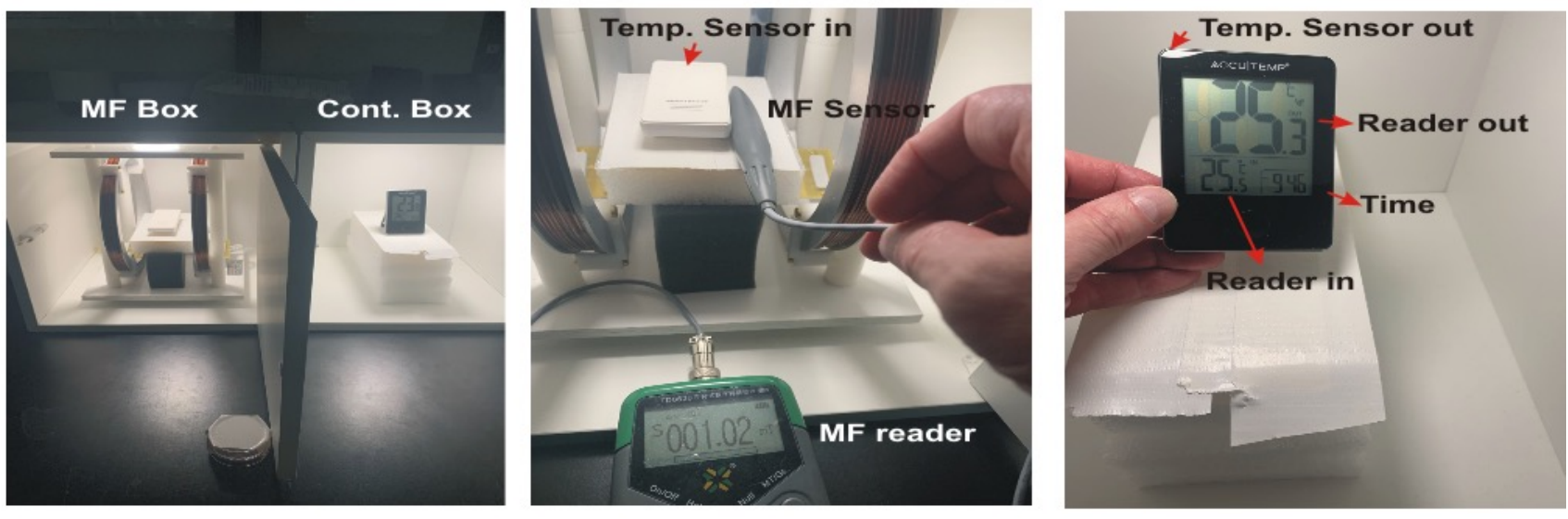}
\caption{Temperature difference between the magnetic field (MF) and control boxes as a function of applied field intensity. Even at the maximum intensity tested (1 mT), the variation remained below 0.2$^{\circ}$C.}
\label{fig:figureS2}
\end{figure}
 
\begin{figure}[H]
\centering
\includegraphics[width=1\textwidth]{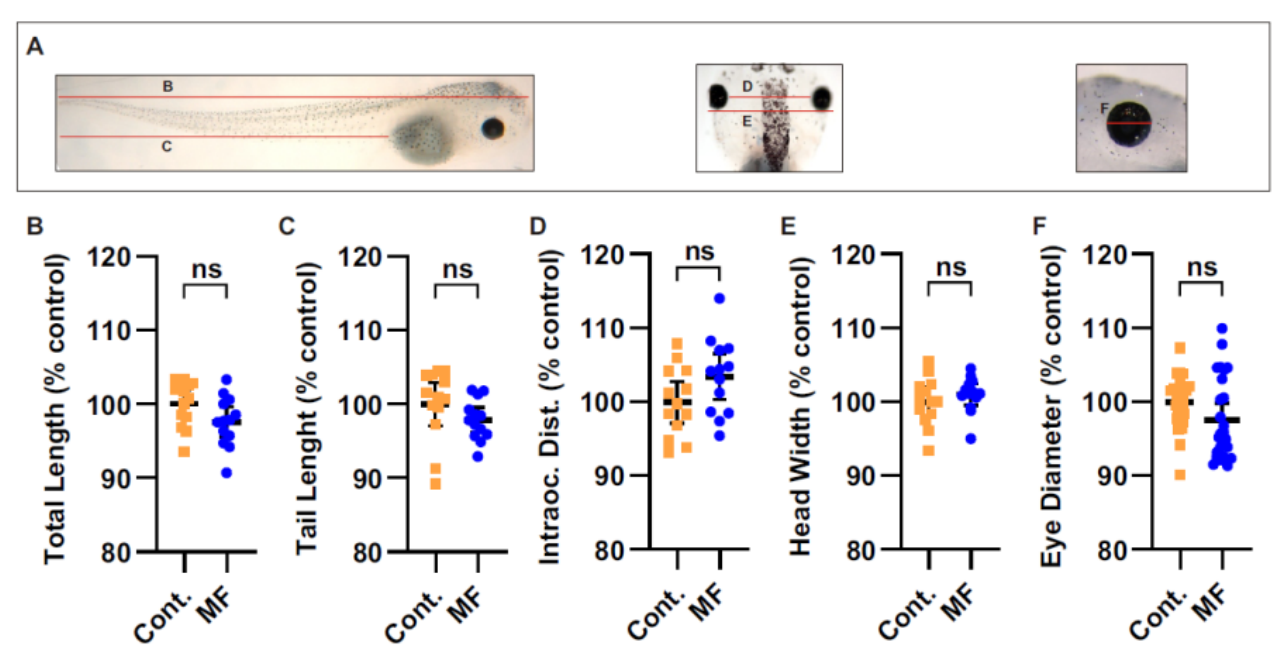}
\caption{\textbf{Magnetic-field exposure does not affect gross morphology.} Morphometric measures (\% of control) in tadpoles exposed to 0.75\,mT from stage~35 for five days versus paired controls. \textbf{(A)}~Features measured; \textbf{(B)}~total length, \textbf{(C)}~tail length, \textbf{(D)}~intraocular distance, \textbf{(E)}~head width, \textbf{(F)}~eye diameter. No measure differed significantly between groups (all $p>0.6$, $n=28$ per group), indicating that magnetic-field exposure does not perturb general development. ns, not significant.}
\label{fig:figureS3}
\end{figure}

\begin{table}[H]
    \centering
    \caption{Isotropic hyperfine coupling constants ($a_\mathrm{iso}$) of the candidate
    radical pairs. For the $\mathrm{FAD}^{\bullet-}$--TrpH$^{\bullet+}$ pair the couplings are fixed at literature values;
    for the generic $(\mathrm{H,H})$--$(\mathrm{H,H})$ pair the four ${}^{1}$H couplings
    are free fit parameters, explored over the indicated range.
    Nuclei: ${}^{14}$N ($I=1$), ${}^{1}$H ($I=\tfrac12$).}
    \begin{tabular}{llcc}
      \toprule
      Radical pair & Radical & Nucleus & $a_\mathrm{iso}$ (mT) \\
      \midrule
      \multirow{4}{*}{FAD$^{\bullet-}$--TrpH$^{\bullet+}$~\cite{Benjamin_2025}}
        & \multirow{2}{*}{FAD$^{\bullet-}$}  & ${}^{14}$N & $+0.570$ \\
        &                                    & ${}^{1}$H  & $+0.420$ \\
        & \multirow{2}{*}{TrpH$^{\bullet+}$} & ${}^{1}$H  & $+1.280$ \\
        &                                    & ${}^{1}$H  & $-0.570$ \\
      
      \midrule
      \multirow{2}{*}{Generic $(\mathrm{H,H})$--$(\mathrm{H,H})$}
        & Radical A & $2\times{}^{1}$H & $|a_\mathrm{iso}|\in[0.4,\,1.0]$ (fit) \\
        & Radical B & $2\times{}^{1}$H & $|a_\mathrm{iso}|\in[0.4,\,1.0]$ (fit) \\
      \bottomrule
    \end{tabular}
    \label{tab:hfcc}
  \end{table}

\begin{figure}[H]
    \centering
    \includegraphics[width=1\textwidth]{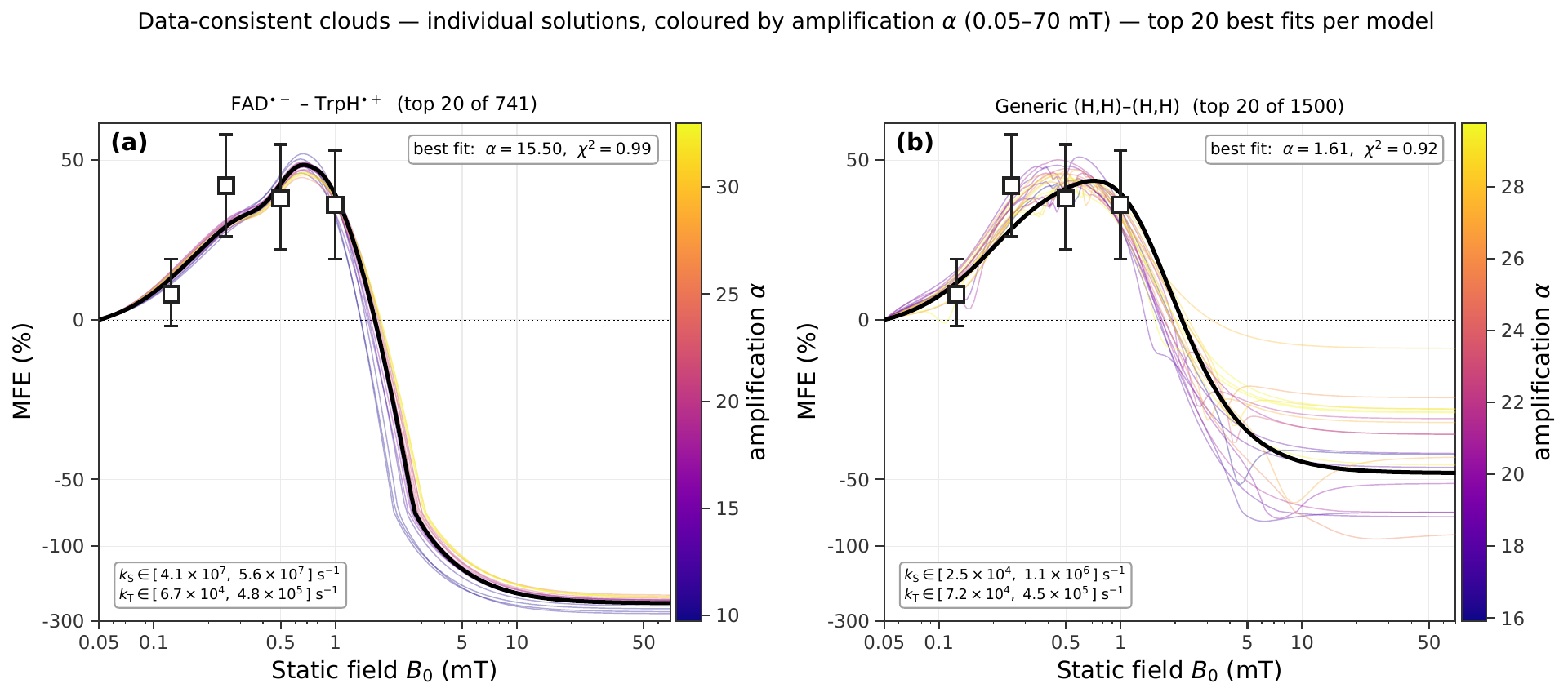}
    \caption{\textbf{Data-consistent solution clouds for the two candidate radical pairs.} For each model, every data-consistent parameter set is shown as an individual magnetic-field-effect curve over $0.05$--$70$\,mT, coloured by its amplification factor $\alpha$ (per-panel colour bar); a set is data-consistent if its predicted effect lies within all four measured error bars. Only the top~20 ($\chi^2$-ranked) solutions per model are drawn; the heavy black curve is the best ($\chi^2$-minimal) fit and the open squares with bars are the experiment (mean $\pm$ measured range). \textbf{(a)}~$\mathrm{FAD}^{\bullet-}$--TrpH$^{\bullet+}$ (top 20 of 741; best fit $\alpha=15.5$, $\chi^2=0.99$; $k_S\in[4.1\times10^{7},5.6\times10^{7}]$\,s$^{-1}$, $k_T\in[6.7\times10^{4},4.8\times10^{5}]$\,s$^{-1}$). \textbf{(b)}~generic $(\mathrm{H,H})$--$(\mathrm{H,H})$ (top 20 of 1500; best fit $\alpha=1.61$, $\chi^2=0.92$; $k_S\in[2.5\times10^{4},1.1\times10^{6}]$\,s$^{-1}$, $k_T\in[7.2\times10^{4},4.5\times10^{5}]$\,s$^{-1}$). Both pairs reproduce the data in the measured $0.125$--$1$\,mT window and diverge at higher field, predicting a maximum near $0.5$\,mT and a sign reversal above $\sim$1.5\,mT. Effect definition and kinetics as in Fig.~\ref{fig:figure2}.}
    \label{fig:figureS4}
\end{figure}

\begin{figure}[H]
\centering
\includegraphics[width=1\textwidth]{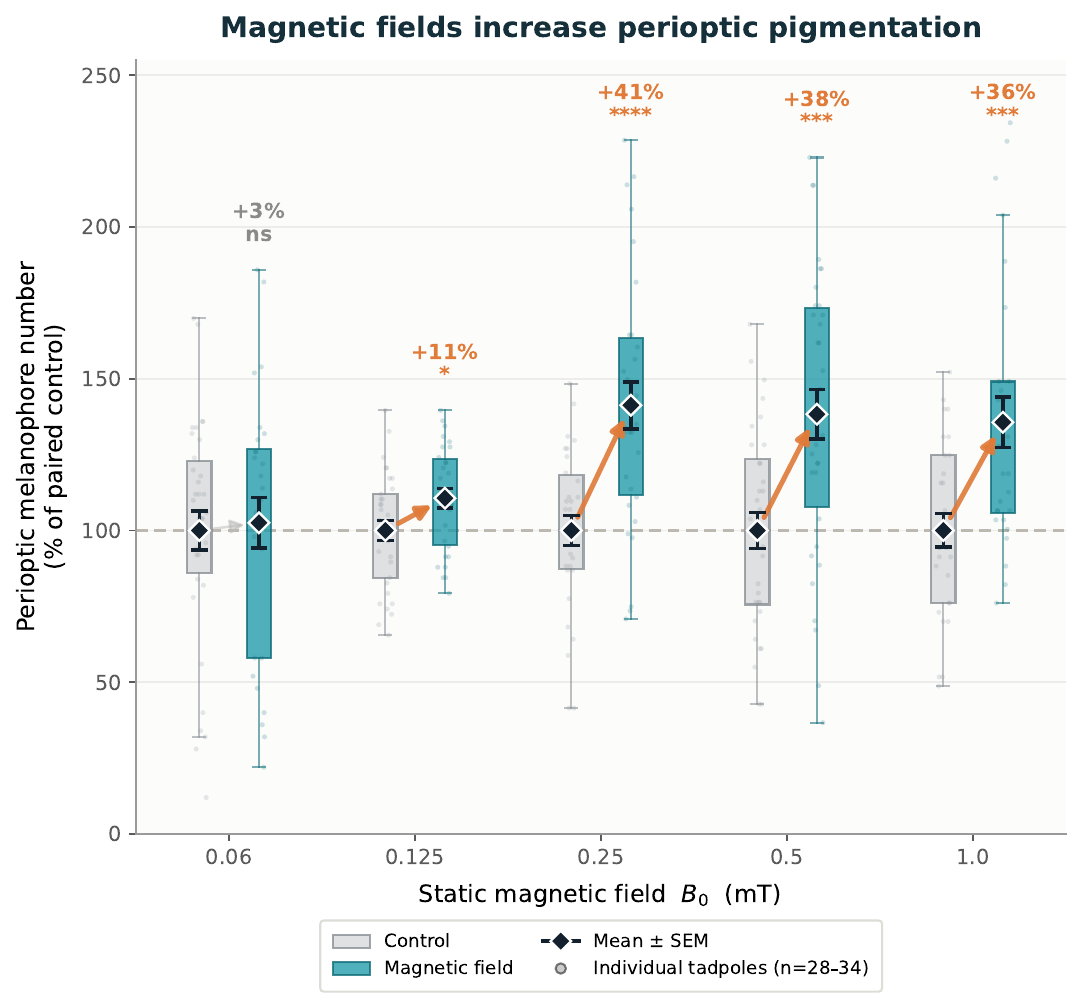}
\caption{\textbf{Magnetic-field effect on perioptic pigmentation (dose--response, day~4).} Perioptic melanophore number (\% of paired control) for control (grey) and magnetic-field-exposed (teal) tadpoles at each static field strength. Boxes, interquartile range; black diamonds, group mean~$\pm$~SEM; faint points, individual tadpoles ($n$~=~28--34 per group); the percentage above each field is the mean magnetic-field effect. Significance versus the paired control by Welch's $t$-test (ns, $^{*}p<0.05$, $^{**}p<0.01$, $^{***}p<0.001$, $^{****}p<0.0001$). These comparisons are uncorrected, in contrast to the ANOVA with Tukey's post-hoc correction used in Fig.~\ref{fig:figure1}; consequently the marginal 0.125\,mT field is flagged significant here but not in the main text.}
\label{fig:figureS5}
\end{figure}
 
\begin{figure}[H]
\centering
\includegraphics[width=1\textwidth]{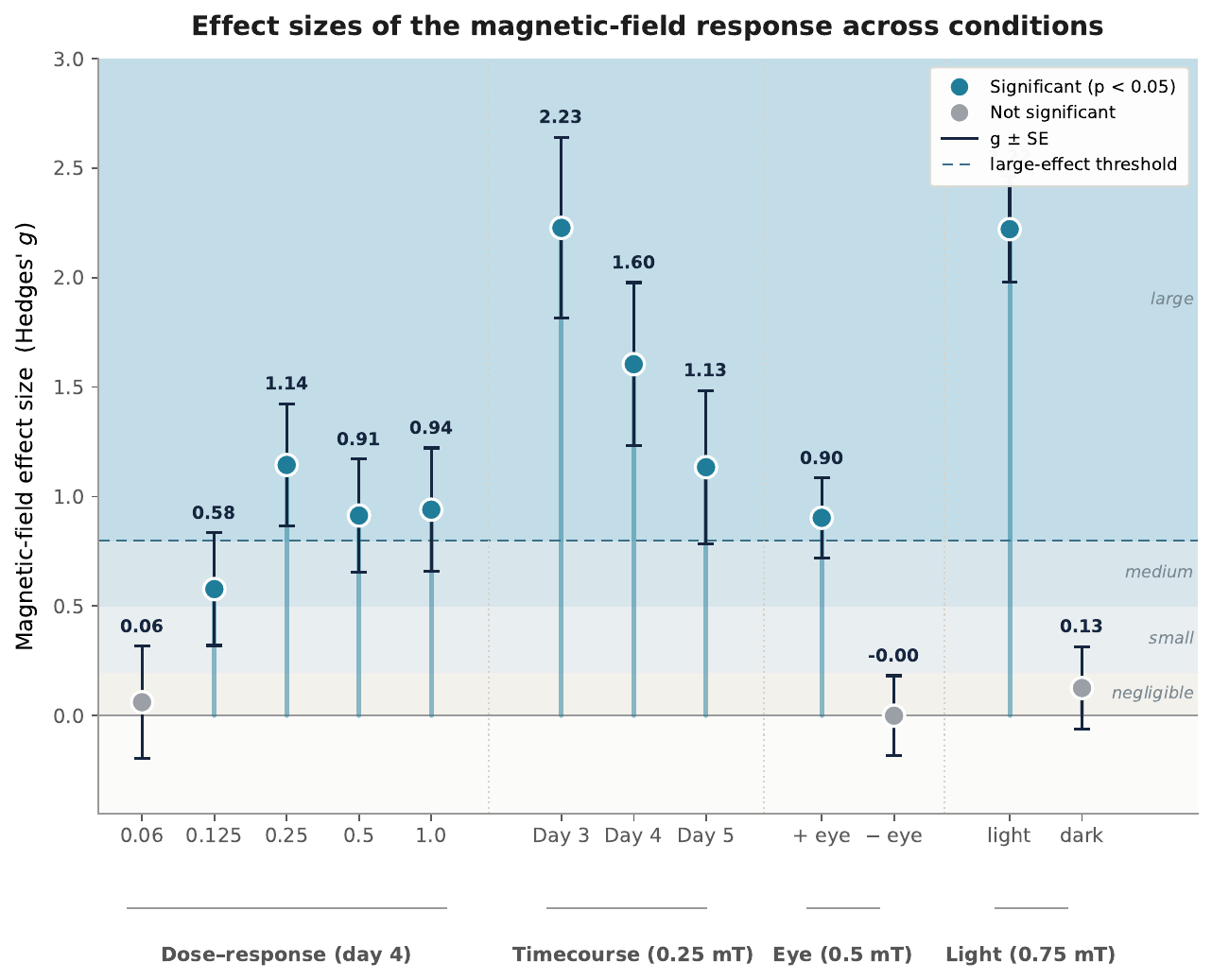}
\caption{\textbf{Standardised effect sizes of the magnetic-field response across conditions.} Hedges' $g$ (magnetic field versus control) with its standard error for the dose--response field strengths (day~4), days 3--5 of the 0.25\,mT timecourse, the enucleation experiment (0.5\,mT; with and without the eye), and the light/dark experiment (0.75\,mT; light/dark cycle versus constant darkness). Shaded bands denote conventional effect-size ranges (negligible/small/medium/large) and the dashed line marks the large-effect threshold ($g=0.8$); points are coloured by statistical significance ($p<0.05$, Welch's $t$-test, uncorrected; cf.\ Supplementary Fig.~\ref{fig:figureS5}). The effect is large where present ($g\approx0.9$--$2.2$) but is abolished without the eye ($g\approx0$) and in constant darkness ($g\approx0.1$).}
\label{fig:figureS6}
\end{figure}

\end{document}